\documentclass[aps,prb,twocolumn,a4,showpacs,superscriptaddress]{revtex4-1} 
\usepackage{amsmath,amssymb,amsfonts,upgreek}
\usepackage{color}
\usepackage{graphicx}
\usepackage{setspace}
\usepackage{multirow}
\usepackage{hhline}
\usepackage{bm}
\usepackage{comment}
\usepackage{natbib}

\begin{document}

\title{Biquadratic and ring exchange interactions in orthorhombic perovskite manganites}

\author{Natalya S. Fedorova}
\email{natalya.fedorova@mat.ethz.ch}
\affiliation{Materials Theory, ETH Z\"{u}rich, Wolfgang-Pauli-Strasse 27, 8093 Z\"{u}rich, Switzerland}
\author{Claude Ederer}
\affiliation{Materials Theory, ETH Z\"{u}rich, Wolfgang-Pauli-Strasse 27, 8093 Z\"{u}rich, Switzerland}
\author{Nicola A. Spaldin}
\affiliation{Materials Theory, ETH Z\"{u}rich, Wolfgang-Pauli-Strasse 27, 8093 Z\"{u}rich, Switzerland}
\author{Andrea Scaramucci}
\email{andrea.scaramucci@psi.ch}
\affiliation{Materials Theory, ETH Z\"{u}rich, Wolfgang-Pauli-Strasse 27, 8093 Z\"{u}rich, Switzerland}
\affiliation{Laboratory for Developments and Methods, Paul Scherrer Institut, 5232 Villigen PSI, Switzerland}

\date{\today}

\begin{abstract}
We use \textit{ab initio} electronic structure calculations within the generalized gradient approximation (GGA+U) to density functional theory (DFT) to determine the microscopic exchange interactions in the series of orthorhombic rare-earth manganites, o-$R$MnO$_3$. Our motivation is to construct a model Hamiltonian (excluding effects due to spin-orbit coupling), which can provide an accurate description of the magnetism in these materials. First, we consider TbMnO$_3$, which exhibits a spiral magnetic order at low temperatures. We map the exchange couplings in this compound onto a Heisenberg Hamiltonian and observe a clear deviation from the Heisenberg-like behavior. We consider first the coupling between magnetic and orbital degrees of freedom as a potential source of non-Heisenberg behavior in TbMnO$_3$, but conclude that it does not explain the observed deviation. We find that higher order magnetic interactions (biquadratic and four-spin ring couplings) should be taken into account for a proper treatment of the magnetism in TbMnO$_3$ as well as in the other representatives of the o-$R$MnO$_3$ series with small radii of the $R$ cation.
\end{abstract}

\maketitle

\section{Introduction}
Perovskite manganites, $R$MnO$_3$ ($R^{3+}$ = rare earth cation), show a great variety of structural, magnetic and electronic phases whose coexistence and interplay give rise to the large diversity of their physical properties. Orthorhombic $R$MnO$_3$ (o-$R$MnO$_3$) exhibiting frustrated magnetic orderings are of particular interest as they belong to the family of so-called magnetoelectric multiferroics - materials, where magnetic and ferroelectric orders are simultaneously presented \cite{{Hill:2000fx},{Spaldin:2005jx},{Cheong:2007uw}}. Indeed, it has been shown experimentally, that the establishment of a spiral ordering of Mn$^{3+}$ spins in TbMnO$_3$ and DyMnO$_3$ is accompanied by the appearance of a spontaneous electric polarization which can be manipulated by an applied magnetic field \cite{Kimura:2003wn}. Recently a magnetically induced electric polarization was also observed in o-HoMnO$_3$ \cite{Lorenz:2007bq}, which has an E-type antiferromagnetic order (E-AFM) \cite{PhysRev.100.545}. Despite the fact that these effects occur at quite low temperatures, the understanding of their mechanisms is important for the fundamental physics of magnetoelectric phenomena and for potential development of multifunctional devices. 

In this work we address the question of the origin of the frustrated magnetic orderings which cause the multiferroic properties in o-$R$MnO$_3$. According to experiment, the magnetic structure in the series of o-$R$MnO$_3$ evolves from A-AFM to the spiral and then to the E-AFM state with decreasing radius of the $R$ cation, which favors the enhancement of orthorhombic distortion. This in turn changes the relative strength of nearest-neighbor (NN) and further neighbor exchange interactions between Mn spins in these materials \cite{{Zhou:2006hj},{Kimura:2003kq}}. This evolution of the magnetic order is usually described within the framework of a Heisenberg model with competing NN and next-nearest-neighbor (NNN) exchanges. Indeed, qualitatively, this model gives the spiral as a ground state for a certain ratio between NN and NNN couplings \cite{{Cheong:2007uw}, {Mochizuki:2009br}, {Kaplan:2009fx}}. However, as we will show in details in Sec.\ \ref{sec:2b}, application of this model for quantitative description of the exchanges in o-$R$MnO$_3$ gives contradictory results. Moreover, it was shown recently, that the E-AFM state cannot  be obtained from the Heisenberg Hamiltonian \cite{Kaplan:2009fx, Kaplan:pRvO8IfS}.
\begin{figure}[h]
\centering
{\includegraphics[scale=0.46, trim=0cm 0cm 0cm 0cm]{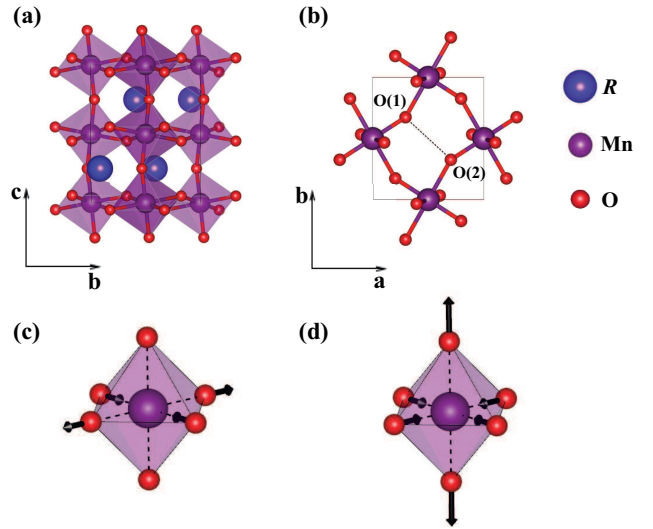}}
\caption
{(Color online) Crystal structure of o-$R$MnO$_3$: (a) - side view; (b) - top view ($R$ ions are not shown). (c) and (d) represent the normal modes of Jahn-Teller distortion $Q_2$ and $Q_3$, respectively.}  
\label{fig1}
\end{figure}

Here we present the results of our studies of the microscopic magnetic couplings in the series of o-$R$MnO$_3$ using first-principles electronic structure calculations with the goal of finding a model Hamiltonian (excluding effects due to spin-orbit coupling) which can accurately describe the magnetism in these materials. First we consider TbMnO$_3$ with spiral spin ordering. We map the exchanges in this compound onto the Heisenberg model and find a clear deviation from Heisenberg-like behavior. We investigate the extent to which this deviation originates from the presence of the orbital ordering in TbMnO$_3$ and show that the coupling between magnetic and orbital degrees of freedom cannot provide the observed deviation. Then we explore the effect of exchange couplings of higher orders than the bilinear exchange (biquadratic and four-spin ring couplings), which are usually neglected. We demonstrate that the higher order contributions are significant in TbMnO$_3$ and other o-$R$MnO$_3$ with small radii of the $R$ cation and they have to be included in the model Hamiltonian for an accurate description of the magnetic properties of orthorhombic manganites. 

This article is organized as follows: in Sec.\ \ref{sec2} we describe the crystal structure, the orbital ordering and its relation to the magnetic properties in o-$R$MnO$_3$, and explain the motivation of our research. Here we also introduce the methods which we use in our calculations and specify the computational details. In Sec.\ \ref{sec3} we calculate the microscopic exchange couplings in TbMnO$_3$ and show that they cannot be described by the Heisenberg Hamiltonian. In Sec.\ \ref{sec4} we discuss the possible sources of the non-Heisenberg behavior in TbMnO$_3$; in particular, we investigate the effects of orbital ordering, structural distortions and higher order exchange couplings. In Sec.\ \ref{sec5} we extend our analysis on the other representatives of the o-$R$MnO$_3$ series, namely, PrMnO$_3$ and LuMnO$_3$. Finally, in Sec.\ \ref{sec6} we summarize our work and give a conclusion.

\section{Motivation, theoretical background and methods}
\label{sec2}
\subsection{Jahn-Teller and GdFeO$_3$-type distortions in o-$R$MnO$_3$}
\label{sec2a}
The o-$R$MnO$_3$ have an orthorhombically distorted perovskite structure (see Fig.\ \ref{fig1}) with space group \textit{Pbnm} ($\#$62) and 20 atoms per unit cell \cite{{Kimura:2003wn},{Alonso:2000kv},{Ederer:2007fc},{Kovacik:2010bf}}. The deviation from the perfect cubic perovskite structure includes the Jahn-Teller distortion of the MnO$_6$ octahedra \cite{Kanamori:1960ki}, their cooperative tiltings \cite{Woodward:1997ci} (the so-called GdFeO$_3$-type, GFO, distortion) and small antiferroelectric displacements of $R$ cations from their ideal positions \cite{Benedek:2013jl}. While the latter structural distortion has been shown to influence the ferroelectric properties, it's  effect on the magnetism is negligible and we do not consider it in this work.

In o-$R$MnO$_3$ each Mn$^{3+}$ ion resides in the middle of an oxygen octahedron with four electrons in $3d$ levels. The crystal field of the perfect octahedron splits the fivefold degenerate $d$ levels into triply-degenerate $t_{2g}$ lower-energy levels and doubly-degenerate $e_g$ levels with higher energy. Electrons occupy the orbitals according to Hund's rules and the Pauli principle, which leads to full occupation of the spin majority $t_{2g}$ states and single occupation of the spin majority $e_g$ states.  Lowering of the symmetry of the crystal field due to the Jahn-Teller effect \cite{Jahn:1937gp} lifts the degeneracy of the $e_g$ electronic state and favors the occupation of a certain orbital which can be represented as a superposition of $d_{z^2}$ and $d_{x^2-y^2}$-states \cite{Kugel:1973ub}:
\begin{equation}
|\psi\rangle=\cos\left(\frac{\theta}{2}\right)|d_{z^2}\rangle+\sin\left(\frac{\theta}{2}\right)|d_{x^2-y^2}\rangle
\label{orbmixing}
\end{equation}
The state $|\psi \rangle$ is uniquely defined by the angle $\theta$ which is called the orbital mixing angle. The corresponding distortion of the octahedron can be written as a linear combination of two normal Jahn-Teller modes $Q_2$ and $Q_3$ \cite{{VanVleck:1939hw},{Kanamori:1960ki}} (Fig.\ \ref{fig1} (c) and (d), respectively): 
\begin{equation}
Q=Q_3\cos\varphi+Q_2\sin\varphi
\end{equation}
The value of $\varphi$ can be estimated with the simple formula
\begin{equation}
\varphi=\arctan\left(\frac{Q_2}{Q_3}\right)=\arctan\left(\frac{\sqrt{3}(l-s)}{2m-l-s}\right),
\label{eq5}
\end{equation}
where  $l$, $m$ and $s$ are the lengths of the long, medium and short Mn-O bonds in the octahedron \cite{Khomskii:2005ws}. The ground state value of $\theta$ is determined by the balance between the energy gain due to the orbital-lattice interaction and the elastic energy cost  \cite{Brink:2004gp}. For a single octahedron this occurs at $\varphi=\theta$. 

Since the oxygen octahedra are interconnected, their distortions and, therefore, the occupied orbital states on neighboring Mn ions are not independent. Below a certain temperature, this leads to a long-range orbital ordering with the orbital mixing angles for two neighboring Mn sites $i$ and $j$ in the $ab$ plane related by: $\theta_i=-\theta_j$ (antiferro-orbital orientation). For nearest neighbors along the $c$ direction they are equal (ferro-orbital orientation).   

The GFO distortion is characterized by almost rigid cooperative rotations of the MnO$_6$-octahedra, which result in the reduction of Mn-O-Mn bond angles and  O(1)-O(2) distances (see Fig.\ \ref{fig1}(b)). In the series of o-$R$MnO$_3$ this distortion increases with decreasing radius of the $R$ cation from La to Lu.
\subsection{\label{sec:2b} Frustrated magnetism in o-$R$MnO$_3$} 
The combination of Jahn-Teller and GFO distortions in o-$R$MnO$_3$ determines their magnetic properties \cite{Lee:2013jl}. According to the Goodenough-Kanamori-Anderson rules \cite{{Goodenoughbook},{Goodenough:1955tg},{Kanamori:1959wa}}, the presence of the orbital ordering of the type which was described in the previous section favors FM exchange coupling between neighboring Mn spins in the $ab$ planes and AFM coupling along the crystallographic $c$ direction. This promotes the establishment of A-AFM ordering  for $R$=La...Gd. However, further decreasing the size of $R$-cation in the series of o-$R$MnO$_3$ (and, therefore, increasing the GFO distortion) causes the transition to the spiral ($R$=Tb, Dy) and then E-AFM states ($R$=Ho...Lu). One can consider the change in the relative strength of FM NN and AFM NNN couplings in the $ab$ plane as an origin of this transition. Indeed, increasing GFO distortion decreases NN exchange as it strongly depends on the Mn-O-Mn bond angles. On the other hand, it enhances the AFM exchange between NNNs along the $b$ axis through the path Mn-O(1)-O(2)-Mn due to the reduction of O(1)-O(2) distances (see Fig.\ \ref{fig1} (b)). This strong AFM NNN exchange causes magnetic frustration \cite{Kimura:2003kq}.

The simplest microscopic model which is often used to discuss this evolution of the magnetic order is the Heisenberg model:
\begin{equation}
H_\mathrm{Heis}=\sum_{\langle i,j\rangle}{J_{ij}\mathbf{S_i}\cdot \mathbf{S_j},}
\label{eqHeis}
\end{equation}
where $J_{ij}$ indicates exchange interactions between spins $\mathbf{S_i}$ and $\mathbf{S_j}$. As exchange interactions are short-ranged, usually only the couplings between first- and second-nearest neighbors are taken into account \cite{Cheong:2007uw}. This model qualitatively explains the establishment of spiral magnetic ordering. Indeed, for o-$R$MnO$_3$ the ratio 
\begin{equation}
\frac{J_{b}}{|J_{ab}|}>\frac{1}{2}
\label{eqSpiral}
\end{equation}
for $ab$ plane FM NN exchange $J_{ab}$ and AFM NNN exchange $J_b$ along the axis $b$ gives a spiral (with a propagation vector along the $b$ axis) as a magnetic ground state \footnote{1/2 appears in Eq.\ \ref{eqSpiral} instead of 1/4 for the case of an infinite spin chain as we take into account the number of equivalent bonds with exchange couplings $J_{b}$ and $J_{ab}$, which are equal to 2 and 4, respectively}. However, the source of E-AFM ordering is still under debate. For example, Kimura \textit{et al.} \cite{Kimura:2003kq} stated that the two-dimensional Heisenberg model with FM NN and certain competing AFM NNN couplings in the $ab$ plane can give E-AFM ordering, whereas Kaplan \cite{Kaplan:2009fx,Kaplan:pRvO8IfS} demonstrated that this state cannot occur in this model unless the biquadratic exchange interaction of the form
\begin{equation}
H_{bq}=\sum_{\langle k,l\rangle} j(\mathbf{S}_k\cdot \mathbf{S}_{l})^2
\label{eqBq}
\end{equation}
is included in the Hamiltonian.  In turn, Solovyev \cite{Solovyev:2009by} claimed that it is crucial to consider the exchange interaction between the third nearest neighbors in the $ab$ planes to stabilize the E-AFM state.

 \begin{figure}
\centering
{\includegraphics[scale=0.47, trim=1.5cm 0.3cm 0.5cm 0.3cm]{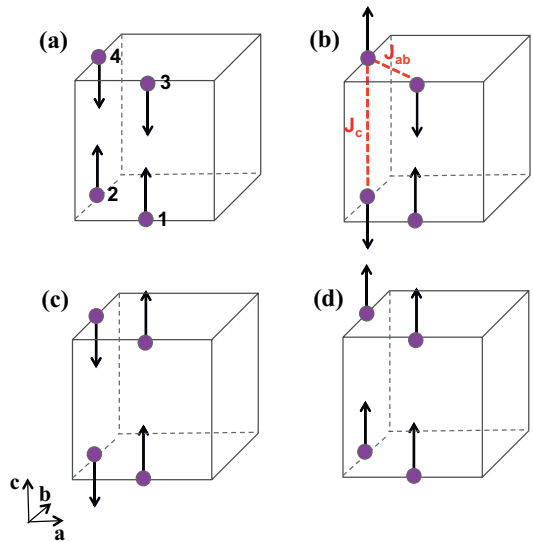}}
\caption
{(Color online) Collinear magnetic orderings (in Wollan-Koehler notation \cite{PhysRev.100.545}) of the Mn spins within the perovskite unit cell: (a) A-AFM, (b) G-AFM, (c) C-AFM and (d) FM. $J_c$ and $J_{ab}$ indicate the NN exchange couplings along the $c$ axis and within the $ab$ planes, respectively.}  
\label{fig2}
\end{figure}

Aside from the disagreement on the source of the E-AFM order, the application of the Heisenberg model for a quantitative description of the magnetism in o-$R$MnO$_3$ gives ambiguous results. Assuming that the magnetism in o-$R$MnO$_3$ is fully described by the Heisenberg Hamiltonian and considering only the couplings between NN spins, the total energy can be written as
\begin{equation}
E=\sum_{\langle i,j\rangle}{J_{ij}\mathbf{S_i}\cdot \mathbf{S_j}+E_0,}
\end{equation}
where $E_0$ includes all other (nonmagnetic) interactions. 
\begin{table}[b]
\caption{The exchange parameters $J_{ab}$ and $J_c$ (in meV) calculated with different theoretical approaches for LaMnO$_3$ using normalized values of spins $\mid \mathbf{S_{i}} \mid$=$\mid \mathbf{S_{j}} \mid$=1.}
\begin{tabular}{p{64pt}|p{40pt}p{40pt}|p{40pt}p{40pt}}
\hline
\hline
& \multicolumn{2}{p{84pt}|}{\centering{$J_{ab}$}} & \multicolumn{2}{p{84pt}}{\centering{$J_{c}$}} \\
\multirow{-2}{60pt}{\centering{Method}} &  \centering{$\Delta E_{AG}$} & \centering{$\Delta E_{FC}$} & \centering{$\Delta E_{FA}$} & \centering{$\Delta E_{CG}$} \tabularnewline
\hline
\centering{GGA-PBE \cite{Yamauchi:2008ia}} & \centering{-27.7} & \centering{-22.5} & \centering{0.5} & \centering{-10.0}  \tabularnewline
\centering{HFA \cite{Solovyev:2009by}} & \centering{-4.75} & \centering{-1.25} & \centering{10.0} & \centering{3.0}  \tabularnewline
\centering{GGA-PW91 \cite{Evarestov:2005ji}} & \centering{-18.0} & \centering{-14.6} & \centering{5.0} & \centering{-1.75}  \tabularnewline
\hline
\hline
\end{tabular}
\label{table1}
\end{table} 
Therefore, one can see that the difference in the total energies of the unit cell of o-$R$MnO$_3$ with A-AFM and G-AFM orientations (see Fig.\ \ref{fig2}(a) and (b), respectively) of Mn$^{3+}$ spins ($\Delta E_{AG}$) defines the value of exchange coupling $J_{ab}$.  Moreover, this value should be the same as that given by the difference in the energies of FM and C-AFM (Fig.\ \ref{fig2}(d) and (c), respectively) states ($\Delta E_{FC}$). Similarly, the NN exchange $J_c$ along the $c$ axis can be extracted from the following energy differences: $\Delta E_{FA}=$ $E$(FM)-$E$(A-AFM) and $\Delta E_{CG}=$ $E$(C-AFM)-$E$(G-AFM) and the obtained values should be the same for these two cases. The energies of FM, A-AFM, C-AFM and G-AFM states have been calculated by several groups for the series of o-$R$MnO$_3$ applying different theoretical approaches \cite{{Yamauchi:2008ia}, {Solovyev:2009by}, {Evarestov:2005ji}}. Using the published values of these energies, we calculate for each case $\Delta E_{AG}$ and $\Delta E_{FC}$, which define $J_{ab}$ and should give the same results. However, we find that the obtained values $\Delta E$ are significantly different. The same is found for $J_c$, where $\Delta E_{FA}$ and $\Delta E_{CG}$ give in some cases even different signs. As an example, we present in Table \ref{table1} the values of $J_{ab}$ and $J_c$ in LaMnO$_3$ obtained using generalized gradient approximation in the form of Perdew, Burke and Ernzerhof (GGA-PBE)\cite{Yamauchi:2008ia}, Hartree-Fock approximation (HFA) \cite{Solovyev:2009by} and GGA with the Perdew-Wang-91 functional (GGA-PW91) \cite{Evarestov:2005ji}. We would like to point out, that we do not compare the values of $J_c$ and $J_{ab}$ obtained with different approximations and presented in different rows of the Table \ref{table1}. For each approximation we compare two values of $J_{ab}$ ($J_c$), which were obtained using $\Delta E_{AG}$ and $\Delta E_{FC}$ ($\Delta E_{FA}$ and $\Delta E_{CG}$) and, in principle, should give very similar values. Table \ref{table1} also demonstrates, that the inconsistencies in the values of exchanges are not related to the choice of the exchange-correlation potential. It should be noted, that the addition of the biquadratic term (Eq. \ref{eqBq}) in the Hamiltonian cannot explain these results as it cancels out in each energy difference. 

These inconsistencies have not been addressed in the literature and require further investigation. On one hand, they could arise from the presence of strong NNN couplings. On the other hand, they could point to the presence of other significant  couplings beyond the Heisenberg Hamiltonian, which have to be taken into account for a proper theoretical analysis of the magnetism in o-$R$MnO$_3$. 

\subsection{Mapping of DFT onto the Heisenberg model}
\label{sec2C}
In order to examine the relevance of the Heisenberg model for o-$R$MnO$_3$ and to clarify the inconsistencies in the previous theoretical results, we perform a thorough analysis of the microscopic exchange couplings by mapping the results of density functional theory (DFT) \cite{Hohenberg:1964ut,Kohn:1965ui} calculations onto the Heisenberg Hamiltonian. For that purpose we use two approaches, described below in this section and based on certain modifications of the initial magnetic states. In this context we point out, that in the DFT calculations periodic boundary conditions are applied and a variation of a state of a Mn spin on one site leads to the same variation of Mn spin states on all periodically equivalent sites.

\subsubsection{Calculations with collinear spin configurations}  
\label{sec2c1}
The first approach, described in detail in Refs. \onlinecite{{Xiang:2011cn},{Whangbo:2005eg}}, is based on calculations of the total energy of the system with collinear spin alignment when the spin states on two sites (let us denote them as 1 and 2) within the given unit cell are modified. If the magnetism in the system is fully described by the Heisenberg Hamiltonian, the energy for such a spin pair can be written as follows: 
\begin{equation}
E=nJ_{12}\mathbf S_1\cdot\mathbf S_2+\mathbf S_1\cdot\mathbf h_1+\mathbf S_2\cdot\mathbf h_2+E_{all}+E_0,
\label{eq_MappDFT}
\end{equation}
where $n$ is the number of equivalent bonds with exchange coupling $J_{12}$, which connect ions 1 and 2 (necessary to take into account the periodic boundary conditions), $\mathbf h_1=\sum_{i\ne 1,2}J_{1i}\mathbf S_i,$  $\mathbf h_2=\sum_{i\ne 1,2}J_{2i}\mathbf S_i$ and $E_{all}=\sum_{i,j\ne 1,2}J_{ij}\mathbf S_i\cdot \mathbf S_j$. The first term in Eq.\ \ref{eq_MappDFT} describes Heisenberg exchange interactions between spins in the considered pair, the second (third) term corresponds to the coupling of the spin 1 (2) with all other spins in the unit cell except spin 2 (1), $E_{all}$ characterizes the exchange couplings between all spins in the unit cell apart from spins 1 and 2, and $E_0$ contains other (nonmagnetic) energy contributions. It is important to use a reasonably large supercell to include in the analysis all couplings which could be significant in a considered system. Four different collinear configurations of the spins 1 and 2 are possible - up-up, up-down, down-up and down-down and their energies can be calculated using DFT. Then, the exchange interaction between these spins can be found using the formula: 
\begin{equation}
J_{12}=\frac{E_{\uparrow\uparrow}+E_{\downarrow\downarrow}-E_{\uparrow\downarrow}-E_{\downarrow\uparrow}}{4nS^2}.
\label{j12}
\end{equation}
For a more direct comparison with other materials we prefer not to normalize our reported values of $J$ by $S^2$ (thus we set $S=1$ for Mn). Substituting in this expression the energies $E_{\uparrow\uparrow}$, $E_{\downarrow\downarrow}$, $E_{\uparrow\downarrow}$ and $E_{\downarrow\uparrow}$ using Eq.\ 8, one sees that all terms, except those describing the exchange interaction between spins 1 and 2, cancel out. As a result, the parameter $J_{12}$ should not depend on the orientation of the spins of the remaining ions in the unit cell. 

\subsubsection{Noncollinear calculations}
\label{sec2c2}
This approach is based on the calculation of the total energy of the system when some spins are rotated away from an initial collinear state \cite{Novak:2008di} and can be illustrated by the example of the unit cell with 4 magnetic ions.
We consider A-AFM ordering for spins in the unit cell as a starting point and rotate the spins of ions 2 and 4 (see Fig.\ \ref{fig2}(a)) by an angle $\alpha$ keeping them antiparallel to each other until we reach G-AFM ordering (Fig.\ \ref{fig2} (b)). The energy of the system as a function of $\alpha$ within the Heisenberg model can be written as:
\begin{equation}
E(\alpha)=-4J_cS^2+8J_{ab}S^2\cos\alpha+E_0,
\end{equation}
and can be calculated using DFT. The resulting curve should fit the form 
\begin{equation}
f(\alpha)=A_1+B_1\cos\alpha
\label{eqFit}
\end{equation}
if the Heisenberg model provides an accurate description (independently of the number of considered exchange couplings as periodic boundary conditions are applied) and the fitting parameter $B_1$ should define the exchange coupling constant $J_{ab}=B_1/8S^2$. $J_c$ can be extracted similarly by rotating spins on sites 3 and 4 from G-AFM to C-AFM ordering (from Fig.\ \ref{fig2} (b) to (c), respectively). 

\subsection{Computational details}
We perform spin-polarized electronic structure calculations using the Vienna \textit{Ab initio} Simulation Package (VASP) \cite{Kresse:1996vf} within the projector-augmented plane wave (PAW) method of DFT. We use the GGA+U approximation for the exchange-correlation potential (in the form of Perdew, Burke and Ernzerhof \cite{Perdew:1996ug}) and apply the parameter of on-site Coulomb repulsion for Mn $d$ states of $U$=2 eV. The parameter of the effective on-site exchange (Hund's rule) interaction $J_H$ is always set to zero. We consider only isotropic exchange interactions, thus spin-orbit coupling is not included in our calculations. To eliminate the effects from the ordering of the f-electron moments of rare-earth ions, we use pseudopotentials for $R$ with the $f$ states frozen in the core. The value of the energy cutoff is set to 600 eV. In all calculations the experimental crystal structures \cite{Alonso:2000kv, Okamoto:2008ey, Wdowik:2011kz} are considered if it is not otherwise specified. The structures are kept fixed to isolate the contributions from spin-lattice coupling. 
To construct the set of projected Wannier functions \cite{Ku:2002bf} we use the Wannier90 \cite{Mostofi:2008ff} code and the VASP2WANNIER90 interface \cite{Franchini:2012fl}. 

\section{Deviation from Heisenberg model}
\label{sec3}
 \begin{figure}
\centering
{\includegraphics[scale=0.61, trim=1.5cm 0.3cm 0.5cm 0.3cm]{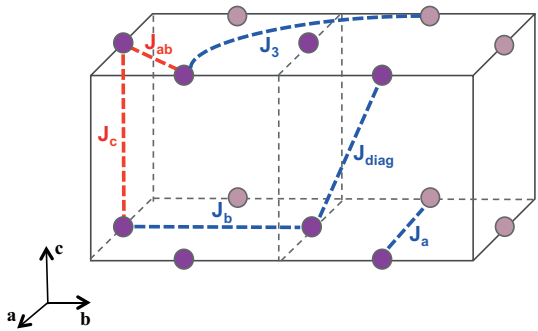}}
\caption
{(Color online) Heisenberg interactions in TbMnO$_3$, which are considered in this work. Mn atoms within the 40 atom supercell are highlighted with dark purple. Light  purple circles indicate Mn ions which belong to the neighboring supercells. NN exchanges are indicated in red, NNN in blue.}  
\label{figEx}
\end{figure}
\begin{table}[b]
\caption{Calculated values of NN and NNN exchange interactions (in meV) in TbMnO$_3$ for FM and A-AFM cases.}
\begin{tabular}{p{40pt}p{30pt}p{30pt}p{30pt}p{30pt}p{30pt}p{30pt}}
\hline
\hline
 & \centering{$J_c$} &\centering{$J_{ab}$} & \centering{$J_a$} & \centering{$J_{diag}$} & \centering{$J_b$} & \centering{$J_3$} \tabularnewline
\hline 
\centering{FM} & \centering{3.68} & \centering{-4.62} & \centering{-0.06} & \centering{0.97} & \centering{1.10} & \centering{1.11} \tabularnewline
\centering{A-AFM} & \centering{-0.85} & \centering{-5.16} & \centering{-0.32} & \centering{-0.10} & \centering{0.68} & \centering{1.26} \tabularnewline
\hline
\hline
\end{tabular}
\label{tab2}
\end{table} 
\begin{table}[b]
\caption{The values of the exchange coupling constant $J_c$ (in meV) in TbMnO$_3$ calculated using the magnetic states shown in Fig.\ \ref{fig40at}. $E_{4sp}$ indicates the contributions to the values of $J_c$ from four-spin ring exchange $K$ between Mn spins confined in adjacent $ab$ planes.}
\begin{tabular}{p{45pt}p{45pt}p{45pt}p{45pt}p{45pt}}
\hline
\hline
 & \centering{(a)} &\centering{(b)} & \centering{(c)} & \centering{(d)} \tabularnewline
\hline 
\centering{$J_c$} & \centering{$-0.88$} & \centering{$-0.68$} & \centering{$0.92$} & \centering{$2.84$}  \tabularnewline
\centering{$E_{4sp}$} & \centering{$-4K$} & \centering{$-4K$} & \centering{0} & \centering{$4K$}  \tabularnewline
\hline
\hline
\end{tabular}
\label{Jc}
\end{table} 
We start with the analysis of the microscopic exchange couplings in the most studied multiferroic orthorhombic perovskite compound TbMnO$_3$ \cite{Xiang:2008jv,Malashevich:2008ew}. We initially assume that the magnetism in this material is fully described by the Heisenberg Hamiltonian (Eq.\ \ref{eqHeis}) and limit ourselves to consideration of the exchange couplings up to third NN within the $ab$ planes and second NN between $ab$ planes. In our notation $J_c$ and $J_{ab}$ are the NN exchanges along the $c$ axis and in the $ab$ plane (see Fig.\ \ref{figEx}) respectively; $J_a$ corresponds to the second NN exchange along the $a$ direction, $J_b$ - along the $b$ axis,  $J_{diag}$ couples second NN in adjacent $ab$ planes; $J_3$ is an exchange between third NN in the $ab$ planes. We extract these parameters applying the method described in Sec.\ \ref{sec2c1}. For this purpose we consider an 80 atom supercell (the orthorhombic unit cell \cite{Alonso:2000kv} is doubled in the $a$ and $b$ directions) and a $\Gamma$-centered  $3\times3\times 5$ k-point mesh. For each $J$ we choose the corresponding spin pair in the supercell and calculate the total energies of the system for the four possible orientations of spins in this pair (up-up, up-down, down-up, down-down). We keep the rest of Mn spins fixed first in the FM state (FM case) and then in the A-AFM state (A-AFM case). The calculated values of $J$ are presented in Table \ref{tab2}.

We find, that $J_b$ is rather weak relative to $J_{ab}$ for both (FM and A-AFM) cases and according to Eq.\ \ref{eqSpiral} cannot produce the spiral state in TbMnO$_3$. The other possible source of frustration could be the AFM coupling $J_3$ which is stronger even than the second NN in-plane couplings $J_a$ and $J_b$. The importance of $J_3$ was already pointed out in Ref.\  \onlinecite{Solovyev:2009by}. Note, that Ref.\ \onlinecite{Solovyev:2009by} proposed strong or weak $J_3$ couplings depending on the relative orientation of the occupied $d$ orbitals on the interacting sites. However, we obtain the same value of $J_3$ for both possible orbital orientations.    

\begin{figure}
\centering
{\includegraphics[scale=0.41]{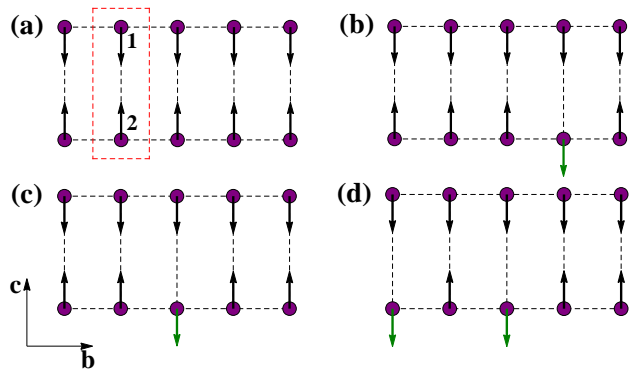}}
\caption
{(Color online) 40 atom supercell of TbMnO$_3$ (side view) with magnetic orders which were used to calculate the exchange parameter $J_c$. Tb and O ions are not shown.}  
\label{fig40at}
\end{figure}
The key result of these calculations is that the values of the exchanges, especially $J_c$, have different magnitudes and in some cases even different signs for FM and A-AFM cases while within the Heisenberg description they should be equal (or at least very similar). 
To double check this result and to determine its origin, we extract $J_c$, which shows the largest inconsistency, with the same method using a 40 atom supercell (20 atom unit cell doubled along the $b$ axis, $7\times 4\times 5$ $\Gamma$-centered k-point mesh). We calculate the total energies switching the direction of spins 1 and 2, but now the rest of spins are kept in the states shown in Fig.\ \ref{fig40at}. The new calculated $J_c$ values are presented in Table \ref{Jc}. One can see that the $J_c$ value obtained using state (a) (which is A-AFM order) is in agreement with the $J_c$ value which was found using the 80 atom supercell and starting from the same magnetic state. Interestingly, the values of $J_c$ are similar for the states (a) and (b), where the closest surroundings of spins 1 and 2 are identical. In turn, if the states differ by the direction of one spin in the nearest neighborhood of the considered spin pair (such as between states (a) and (c) or (c) and (d)), $J_c$ changes by approximately the same amount (in average by 1.85 meV). This suggests the presence of strong couplings beyond the Heisenberg Hamiltonian which involve in some way the magnetic interactions between the nearest-neighboring Mn spins.
\begin{figure}
\centering
{\includegraphics[trim=1cm 0cm 0cm 0cm,scale=0.52]{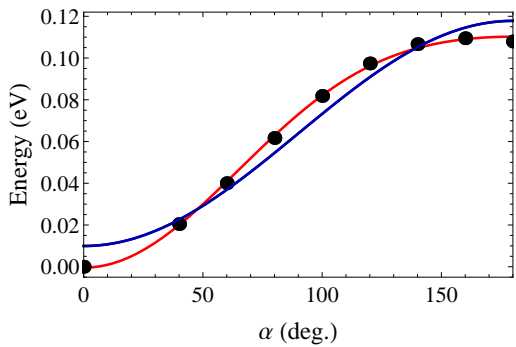}}
\caption
{(Color online) Dependence of the energy $E$ (relative to the energy of A-AFM order) of TbMnO$_3$ on the rotation angle $\alpha$ of spins from A-AFM to G-AFM state. The results of DFT calculations are shown by dots and the fitting to the Heisenberg model (Eq.\ \ref{eqFit}) by the blue line. The red line indicates the fitting to a Hamiltonian which includes bilinear and higher order exchange couplings.}  
\label{fitToHeis}
\end{figure}

\begin{figure}
\includegraphics[scale=0.42,trim=1cm 0cm 0cm 0cm]{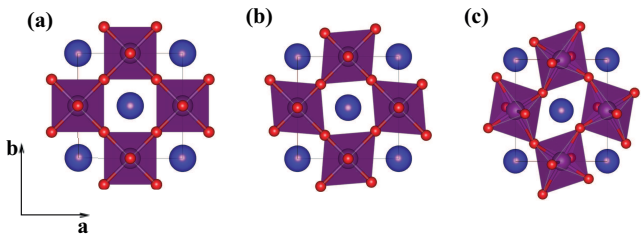}
\caption{(Color online) (a) Ideal cubic perovskite structure; (b) Purely JT distorted structure (tetragonal); (c) Fully JT+GFO distorted structure (orthorhombic) of TbMnO$_3$.}  
\label{figAnalysis}
\end{figure}
Another check can be done by the method described in Sec.\ \ref{sec2c2}. To apply this method we consider a 20 atom unit cell and a $5\times 5\times 3$ k-point mesh. We rotate the spins on sites 2 and 4 from A-AFM to G-AFM ordering (see Fig.\ \ref{fig2}) and calculate the energies $E$ of the system (relative to the energy of A-AFM order) for several values of spin rotation angle $\alpha$ between 0$^{\circ}$ and 180$^{\circ}$. $E(\alpha)$ and its fitting to $f(\alpha)$ (Eq.\ \ref{eqFit}) are presented in Fig.\ \ref{fitToHeis} (black dots and blue line, respectively). One can see that $E(\alpha)$ shows clear deviation from the cosinusoidal behaviour predicted by the Heisenberg model. 

These results lead us to the conclusion that the Heisenberg Hamiltonian in its usual form is not able to accurately describe the magnetism in TbMnO$_3$ and more couplings have to be taken into consideration.

\section{Origin of non-Heisenberg behavior}
\label{sec4}
\subsection{Orbital ordering}
First, we investigate whether the observed non-Heisenberg behavior can originate from the presence of the orbital ordering in TbMnO$_3$. Indeed, as was already described in Sec.\ \ref{sec:2b}, the magnetic and orbital orderings are related as far as the coupling of spins on neighboring Mn$^{3+}$ ions is determined by the occupation of their particular orbitals through the superexchange mechanism. To take this behavior into account, the ordinary superexchange was generalized for the case of systems with orbital degeneracy by Kugel and Khomskii \cite{Kugel:1973ub}. They introduced a model Hamiltonian, in which, besides the Heisenberg exchange, they included terms describing orbital-orbital and orbital-spin couplings. The latter gives the change in the orbital ordering with variation of the spin alignment (or vice versa) and, if it is large enough, can explain the different values of exchange obtained for A-AFM and FM cases as well as the deviation from Heisenberg behavior observed in noncollinear calculations.  

\begin{figure}[t]
\centering
{\includegraphics[scale=0.5]{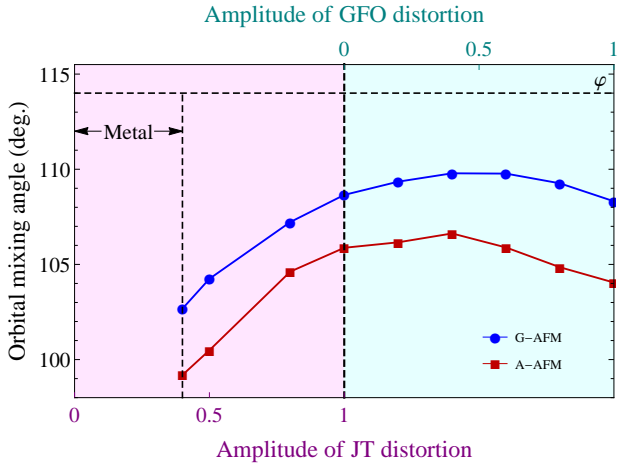}}
\caption
{(Color online) Orbital mixing angle as a function of the amplitudes of JT and GdFeO$_3$-type distortions for G-AFM and A-AFM magnetic orderings in TbMnO$_3$. In the part of the graph highlighted with violet (cyan), only the amplitude of JT (GFO)-distortion is varied. $\varphi$ is determined using Eq.\ \ref{eq5}.}  
\label{mixangles}
\end{figure}

The occupied $e_g$ orbital $|\psi \rangle$ for each Mn$^{3+}$ ion is uniquely defined in terms of the orbital mixing angle $\theta$ by Eq.\ \ref{orbmixing}. 
To extract $\theta$ for TbMnO$_3$, so as to trace its evolution with structural distortions and to estimate the strength of coupling between orbitals and spins we perform the following analysis: We construct a perfect cubic perovskite structure for TbMnO$_3$ using a 20 atom unit cell and keeping the volume of each MnO$_6$-octahedron equal to the experimental one (see Fig.\ \ref{figAnalysis} (a)). Then we start to apply the JT modes Q$_2$ and Q$_3$ (without GFO distortion) in such a way that $Q_{i,applied}=aQ_{i,exp}$ (thus the angles $\varphi$, which are defined by Eq.\ \ref{eq5}, are equal for all values of $a$). $a$ is varied from 0 to 1 and $Q_{i,exp}$ corresponds to the structure with the full JT distortion (Fig.\ \ref{figAnalysis} (b)). To reach a more transparent description of the orbital ordering, we use a representation in terms of Wannier functions \cite{Marzari:2012eu} (WF), which, unlike Bloch functions, are localized in space and have minimal overlap with the surrounding orbitals. To construct a set of $e_g$ like WFs, we proceed similarly to Refs.\ \onlinecite{Yin:2006hd, Ku:2002bf}. We calculate the Bloch functions within GGA+U (using the structures corresponding to different $a$), and for each structure we define an energy window, in which $e_g$ bands are located, based on projected densities of states and band structures. Then, using the VASP2WANNIER90 interface and Wannier90 code, we construct four WFs via projection of atomiclike $|d_{z^2}\rangle$ and $|d_{x^2-y^2}\rangle$ orbitals centered on two Mn sites (1 and 2) on the majority spin Bloch bands within the chosen energy window.  Then we calculate the occupation matrices in the basis of these WFs for several values of $a$ and two types of ordering of the Mn magnetic moments (A- and G-AFM). Solving the eigenvalue problem for these matrices, we find $|\psi_i \rangle$ ($i=1,2$) and, thus, $\theta_i$. We plot $\theta_1$ as a function of $a$ starting from $a$=0.4 (see Fig.\ \ref{mixangles}, left half of the graph), since smaller amplitudes of JT distortion give a metallic ground state. $\theta_2$ has the same values as $\theta_1$, but the opposite sign. 
 
\begin{figure}
\centering
{\includegraphics[scale=0.36, trim=0.6cm 0cm 0cm 0cm]{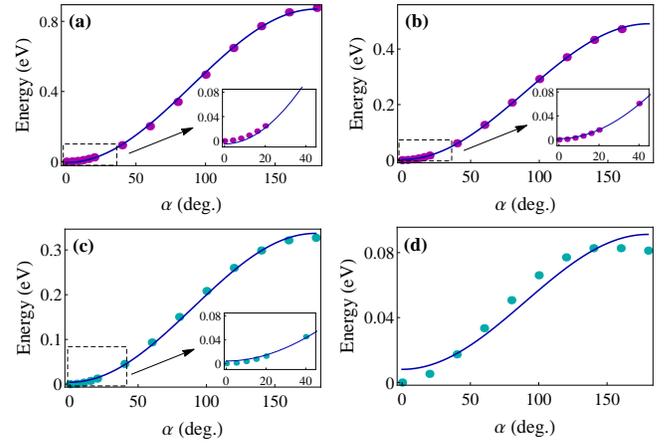}}
\caption
{(Color online) Dependence of the energy $E$ (relative to the energy of A-AFM order) on the rotation angle $\alpha$ of spins from A-AFM to G-AFM state for the structures of TbMnO$_3$ with different amplitudes of JT and GFO-distortions. Dots indicate the results of DFT calculations, lines show the fittings to the Heisenberg Hamiltonian (Eq.\ \ref{eqFit}). (a) and (b) correspond to the structures with 40\% and 100\% JT distortion, respectively, without octahedral tiltings; (c) and (d) to the structures with the full JT distortion and 60\% and 100\% GFO distortion, respectively. Plot (d) was obtained using the crystal structure which unlike the experimental one does not include the antiferroelectric displacements of $R$ cations, thus it is not identical to the one shown on Fig.\ \ref{fitToHeis}.}  
\label{figJTGFO}
\end{figure}

As we expected, the calculated $\theta$ are different for A- and G-AFM orderings. With increasing JT distortion from 40 to 100\% the mixing angles for both AFM orders change by approximately $6^\circ$ and tend to the value of $\varphi\approx 114^\circ$ which is imposed by the structure ($\varphi$ is calculated using Eq.\ \ref{eq5} and experimental lengths of long, medium and short bonds in Mn-O octahedra). It is important, that the difference in the orbital mixing angles  $\Delta \theta$ between A- and G-AFM orderings (in other words, the variation of the orbital ordering by the change in the magnetic structure) is quite small for the whole range of JT distortion amplitudes and reaches a maximum of $\Delta\theta\approx 3^\circ$. In order to check, whether such a small variation of the orbital mixing angle $\Delta\theta$ can cause the deviation from Heisenberg behavior 
which was found in Sec.\ \ref{sec3}, we perform the calculations of the total energies rotating the spins from A-AFM to G-AFM ordering and using the structures with different amplitudes of JT distortion ($a=0.4$ and 1). The obtained angular dependences of the total energy and their fittings to $f(\alpha)$ (Eq.\ \ref{eqFit}) are shown in Fig.\ \ref{figJTGFO} (a), (b). It is clearly seen that the calculated $E(\alpha)$ fit well with the Heisenberg Hamiltonian for both amplitudes of JT distortion. Therefore, one can conclude, that the variation in $\theta$ associated with the change in the magnetic order is not sufficient to explain the large deviation from the Heisenberg model which was observed in our previous calculations. It should be taken into account, however, that the energy scale is 4-6 times larger than in the case where we perform the calculations using the experimental crystal structure (Fig.\ \ref{fitToHeis}). This is because in the latter case the exchange energy is reduced by the presence of GFO distortion. Thus, it is also possible, that the contribution from $\Delta\theta$ is not significant in comparison with the strong exchange energy within the tetragonal structure, but could be important when the orthorhombic distortion comes into play. Therefore, we are motivated to analyse next the effect of GFO distortion on the orbital and magnetic orderings. 

To investigate the variation of the orbital mixing angle by GFO distortion we again construct four projected WFs. In this case, to initialize projections, we introduce a local coordinate system for each MnO$_6$ octahedron in such a way that $x$, $y$ and $z$ axes are aligned as much as possible along the long, short and medium Mn-O bonds, respectively. Other than that, we proceed in the same way as before: Starting from the fully JT distorted structure (Fig.\ \ref{figAnalysis} (b)),  we gradually increase the octahedral rotations to reach the experimentally observed Mn-O-Mn bond angles. The final structure is shown in Fig.\ \ref{figAnalysis} (c); in comparison with the experimental structure, this one does not include a small antiferroelectric shift of Tb cations. We calculate the orbital mixing angles as a function of the amplitude of GFO distortion for A- and G-AFM orderings (Fig.\ \ref{mixangles}, right part of the graph). Then we perform spin rotations from A- to G-AFM ordering with 60\% and 100\% GFO distorted crystal structures. Corresponding angular dependences of the total energy are presented in Fig.\ \ref{figJTGFO} (c), (d).

We find, that increasing GFO distortion causes smaller variation of the orbital mixing angle ($\approx 1.5-3^\circ$) in comparison with JT distortion for both magnetic orderings. Moreover, it almost does not affect $\Delta \theta$ between different types of magnetic ordering (indeed, curves for A-AFM and G-AFM stay almost parallel). However, it induces and enhances the deviation of $E(\alpha)$ from $f(\alpha)$ as shown in Fig.\ \ref{figJTGFO} (c), (d). Therefore we conclude that non-Heisenberg behavior originates from the modification of Mn-O-Mn bond angles due to the reduction of the energy of the exchange interactions between NN Mn spins, which makes weak energy contributions more significant.  

\begin{figure}
\centering
{\includegraphics[scale=0.345, trim=1cm 1cm 2cm 1cm]{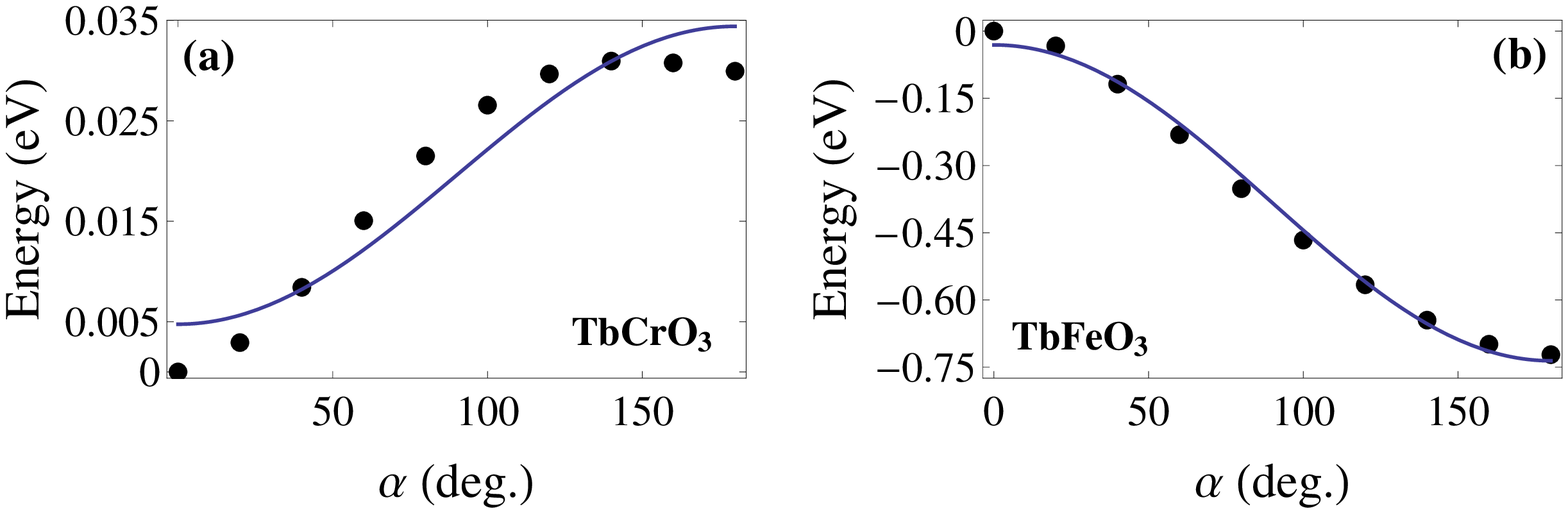}}
\caption
{(Color online) Dependence of the energy $E$ (relative to the energy of A-AFM order) on the rotation angle $\alpha$ of spins from A-AFM to G-AFM state for compounds without orbital ordering: (a) TbCrO$_3$ and (b) TbFeO$_3$. Dots correspond to the results of DFT calculations, lines show the fittings to the Heisenberg Hamiltonian (Eq.\ \ref{eqFit}).}  
\label{figCrFe}
\end{figure}

In order to check whether these weak contributions are provided by $\Delta \theta$, we perform the same spin rotations for two compounds which do not contain JT active ions (therefore, do not have an orbital ordering), TbCrO$_3$ (Cr$^{3+}$: $t^3_{2g}e^0_g$) and TbFeO$_3$ (Fe$^{3+}$: $t^3_{2g}e^2_g$). In these calculations we use the structure of TbMnO$_3$ with Mn$^{3+}$ replaced by Cr$^{3+}$ and Fe$^{3+}$, respectively, which allows us to modify the strength of NN exchange interactions by changing the occupation of $d$ orbitals without any variation of the crystal structure. Indeed, in TbCrO$_3$ magnetic couplings are mostly provided by the hopping processes between $t_{2g}$ orbitals (mediated by oxygen $p$ states) as $e_g$ orbitals are empty, whereas in TbFeO$_3$ both $t_{2g}$ and $e_g$ states of one spin direction are fully occupied and participate in exchange interactions. Moreover, the $e_g$ orbitals in octahedral coordination have stronger overlap with O $p$ states than the $t_{2g}$ due to their geometry, and therefore provide stronger coupling. As a result, one can expect significantly larger magnitudes of NN exchanges for TbFeO$_3$ than for TbCrO$_3$. 

The calculated $E(\alpha)$ are presented in Fig.\ \ref{figCrFe}. Both compounds exhibit deviations of $E(\alpha)$ from the cosinusoidal behavior even in the absence of an orbital ordering. Notably, the deviation is stronger for the case of TbCrO$_3$ than for TbFeO$_3$. We assume that in TbCrO$_3$ the couplings which are not considered in the Heisenberg Hamiltonian are comparable in magnitude with NN exchanges between Mn $t_{2g}$ states and cause a stronger deviation from cosinusoidal behavior, whereas in TbFeO$_3$ they are not significant relative to the strong NN exchange and the Heisenberg model works sufficiently well. Thus, the observed non-Heisenberg behavior cannot be explained by the presence of the orbital ordering. Instead, it appears in the materials where the energy of the exchange couplings is reduced by the modification of Mn-O-Mn bond angles or by the occupation of the orbitals participating in the superexchange.

\subsection{Higher order exchange couplings}
\label{sec4b}
Next we investigate whether exchange couplings of higher order than the usual bilinear term might be responsible for the observed deviation from the Heisenberg model. Generally speaking, the higher order exchanges as well as the bilinear coupling can be derived from a half-filled Hubbard model in the limit $t/U<<1$ (which is applicable for insulators),
\begin{equation}
\label{eq19}
H=-t\sum\limits_{\langle i,j\rangle,\sigma}\left(c_{i\sigma}^{\dagger}c_{j\sigma}+c_{j\sigma}^{\dagger}c_{i\sigma}\right)+U\sum\limits_{j}\hat{n}_{j\uparrow}\hat{n}_{j\downarrow},
\end{equation}
where $t$ is a hopping parameter, $U$ describes the on-site Coulomb repulsion, $c_{j\sigma}^{\dagger}$ and $c_{j\sigma}$ are operators of creation and annihilation of electrons with spin $\sigma$ in the Wannier state $w(\mathbf{r}-\mathbf{R}_j)$ and $\hat{n}_{j\uparrow}=c^\dagger_{j\uparrow}c_{j\uparrow}$ is the occupation number operator.  Second order perturbation theory in $t$ gives the energy correction in the form of Heisenberg exchange, whereas the fourth order gives biquadratic, four-spin ring interactions and additional contributions to NNN couplings. The four-spin ring term describes the consecutive hopping processes between NN ions forming a four-site plaquette and has the following form \cite{{Fazekasbook},{Takahashi:1976vn}}:
\begin{eqnarray}
H_{4sp} & \propto &  \left[ \left(\mathbf S_i\cdot\mathbf S_j\right)\left(\mathbf S_k\cdot\mathbf S_l\right)+  \left(\mathbf S_i\cdot\mathbf S_l\right)\left(\mathbf S_k\cdot\mathbf S_j\right) \right.  \nonumber \\ 
&-&  \left. \left(\mathbf S_i\cdot\mathbf S_k\right)\left(\mathbf S_j\cdot\mathbf S_l\right)\right],
\label{4body}
\end{eqnarray}
where $i$, $j$, $k$ and $l$ enumerate spins of the plaquette.
For the Heisenberg model to be valid, all higher order terms should be negligible compared with the bilinear term.  As their strength is defined by $t^4/U^3$, and that of the bilinear term by $t^2/U$, this should be the case in the limit of small enough $t/U$. However, several theoretical and experimental groups found that in some compounds these terms are significant. For example, it was shown, that the results of a paramagnetic resonance study \cite{Harris:2011wd} of pairs of Mn$^{2+}$ ions in MgO fit much better with a Hamiltonian that includes biquadratic exchange $H_{bq}$ (Eq.\ \ref{eqBq}) than with the ordinary Heisenberg Hamiltonian. Later, the significance of $H_{bq}$ was invoked to explain the establishment of the E-AFM ordering \cite{Kaplan:2009fx} in o-$R$MnO$_3$ with $R$=Ho...Lu as we mentioned in Sec.\ \ref{sec:2b}.  The four-spin ring interaction was found to be important to explain the dispersion of the magnetic excitations in La$_2$CuO$_4$ measured using inelastic neutron scattering experiments \cite{Coldea:2001gu}. It was also shown to be significant in the spin-ladder cuprates SrCu$_2$O$_3$, CaCu$_2$O$_3$ and Sr$_2$CuO$_4$ using \textit{ab initio} quantum chemistry embedded cluster calculations \cite{Calzado:2003iy}. 

First, let us check whether the presence of the higher order terms can explain the inconsistent values of the exchange coupling constant $J_c$ in TbMnO$_3$ which were obtained in Sec.\ \ref{sec3} starting from different states with collinear spin alignment. 
We already mentioned, that the addition of $H_{bq}$ cannot affect the resulting values of exchanges as the applied method considers the energy differences between states with collinear spin orientations and in these differences biquadratic terms always cancel out. Fourth order contributions to NNN interactions, if present, are already included in the analysis as they cannot be distinguished from the bilinear NNN couplings. To introduce the terms describing the four-spin ring exchanges we have to consider the couplings between spins in the plaquettes confined in the $ab$ planes as well as from those that contain pairs of Mn spins from neighboring $ab$ planes. We denote the corresponding coupling constants as $G$ and $K$ (see Fig.\ \ref{bqCalc} (a)). Thus, we can write the energies $E_{\uparrow \uparrow}$, $E_{\uparrow \downarrow}$, $E_{\downarrow \uparrow}$ and $E_{\downarrow \downarrow}$ for the 80 atom supercell of TbMnO$_3$ in A-AFM and FM cases including four-spin interactions and put them in Eq.\ \ref{j12} to extract $J_c$. In this way we find that the in-plane ring exchanges $G$ cancel each other for both cases in the linear combinations of these energies and obtain (for $S=1$)
\begin{equation}
\label{jcwith4b}
\begin{gathered}
J_c\mathrm{(A-AFM)} \to J_c-4K \\
J_c\mathrm{(FM)} \to J_c+4K.
\end{gathered}
\end{equation}
This result shows that the presence of $K$ is the most likely origin of the difference in the obtained values of $J_c$. Using Eqs.\ \ref{jcwith4b} and values of $J_c(\mathrm{A-AFM})$ and $J_c(\mathrm{FM})$ which were obtained in Sec.\ \ref{sec3}, one finds the value of $K\approx 0.6$ meV for TbMnO$_3$.
\begin{figure}
\centering
{\includegraphics[scale=0.4,trim= 0.8cm 0cm 0cm 0cm]{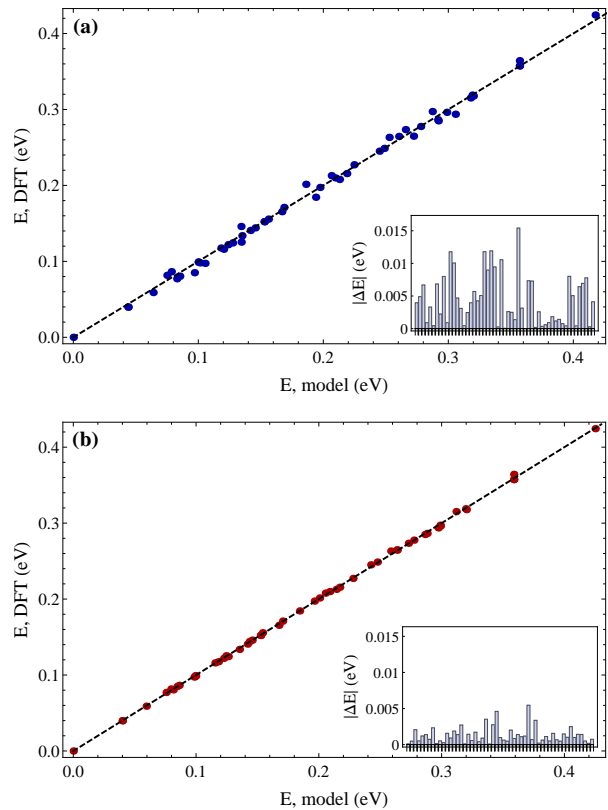}}
\caption
{(Color online) Energies of the 80 atom supercell of TbMnO$_3$ with 54 inequivalent collinear magnetic configurations (referred to the energy of the A-AFM state) predicted by (a) the pure Heisenberg Hamiltonian and (b) the Hamiltonian, which includes bilinear and four-spin ring couplings, and plotted versus the energies of corresponding states calculated using DFT. Ideally, the model and DFT energies should be equal and points should lie on the dashed line. Insets show the deviations of the model energies from those calculated using DFT. Each bar corresponds to one considered magnetic configuration.}  
\label{leastsquares}
\end{figure}
\begin{table*}[!htb]
\caption{Coupling constants (in meV) for bilinear, four-spin ring and biquadratic exchanges in TbMnO$_3$ (for $U$=2 eV and $U$=1 eV), PrMnO$_3$ (for $U$=1eV) and LuMnO$_3$ (for $U$=1 eV).}
\begin{tabular}{p{52pt}p{32pt}p{38pt}p{38pt}p{38pt}p{38pt}p{38pt}p{38pt}p{38pt}p{38pt}p{38pt}p{38pt}}
\hline
\hline
 & \centering{$U_\mathrm{Mn}$} & \centering{$J_c$} & \centering{$J_{ab}$} & \centering{$J_a$} &\centering{$J_{diag}$} & \centering{$J_b$} & \centering{$J_3$} & \centering{$G$} & \centering{$K$} & \centering{$j_c$} & \centering{$j_{ab}$}  
\tabularnewline
\hline 
\centering{TbMnO$_3$} & \centering{2 eV} & \centering{1.22} & \centering{-6.01} & \centering{-0.47} & \centering{0.31} & \centering{0.65} &\centering{1.21} & \centering{-0.05} & \centering{0.50} & \centering{-0.81} & \centering{-2.29}
\tabularnewline
\hline
\centering{PrMnO$_3$} & \centering{1 eV} & \centering{1.79} & \centering{-14.16} & \centering{-0.48} & \centering{0.88} & \centering{0.26} &\centering{3.25} & \centering{-0.07} & \centering{0.80} & \centering{-2.61} & \centering{-2.95} 
\tabularnewline
\centering{TbMnO$_3$} & \centering{1 eV} & \centering{4.26} & \centering{-3.86} & \centering{-0.37} & \centering{0.58} & \centering{0.85} &\centering{1.77} & \centering{-0.02} & \centering{0.77} & \centering{-0.47} & \centering{-2.09}
\tabularnewline
\centering{LuMnO$_3$} & \centering{1 eV} & \centering{3.76} & \centering{-0.48} & \centering{-0.55} & \centering{0.53} & \centering{0.93} &\centering{1.75} & \centering{0.15} & \centering{0.66} & \centering{-0.37} & \centering{-2.29}
\tabularnewline
\hline
\hline
\end{tabular}
\label{allcouplings}
\end{table*} 
The same can be done for the 40 atom supercell of TbMnO$_3$ with magnetic configurations shown in Fig.\ \ref{fig40at}. As before, here we find that the in-plane four-spin couplings $G$ cancel each other. The contributions to $J_c$ arising from the inter-plane ring exchanges obtained for these states are summarized in Table \ref{Jc}. One can see the relation between these contributions and the values of $J_c$ which were calculated with DFT using the structures (a)-(d) and presented in the first line of Table \ref{Jc}. 
Indeed, for the states (a) and (b), the interplane four-spin exchanges contribute exactly the same to $J_c$ ($E_{4sp}=-4K$), and the values of $J_c$ which we extracted using DFT for these states are very similar. States (a) and (c) as well as (c) and (d)  have contributions to $J_c$ which differ by $4K$. Notably, the $J_c$ values which we extracted for these states vary by approximately the same amount (in average $1.85$ meV). This gives the value of $K\approx 0.5$ meV, which is in agreement with the value of $K$ obtained using Eqs.\ \ref{jcwith4b}. Thus we confirm the presence of the strong four-spin interplane exchange couplings in TbMnO$_3$ and show that the addition of these couplings to the model Hamiltonian can explain the inconsistent values of NN exchanges which were found in Sec.\ \ref{sec3}. 

The size of the in-plane four-spin coupling can be estimated similarly by choosing the appropriate collinear spin states and calculating energy differences for them. However, we proceed in a different way. As we already calculated the total energies of the 80 atom supercell of TbMnO$_3$ for a large number (namely 54) of inequivalent magnetic collinear states, we can write the energies of these states using the model Hamiltonian that includes bilinear and four-spin ring couplings and construct an overdetermined system of linear equations, where the unknowns are the exchange coupling constants (bilinear ones (see Fig.\ \ref{figEx}): $J_c$, $J_{ab}$, $J_a$, $J_{diag}$, $J_b$, $J_3$ and four-spin ones (Fig.\ \ref{bqCalc} (a)): $G$ and $K$). To build this system of equations we use only the states which are insulating and take the energy of the A-AFM state as a reference. Then we use the least mean square method to extract all coupling constants. The obtained values are presented in Table \ref{allcouplings}. We find that the in-plane four-spin coupling $G$ is negligible in comparison with the inter-plane one $K$. Further investigation is required to find an explanation for this observation. 

Using our extracted values of the coupling constants, we calculate the expected energies of all 54 states using the considered model Hamiltonian. We plot them versus the energies of these states (referred to the energy of the A-AFM state) calculated using DFT in order to examine how well our model predicts the magnetic properties of the system (ideally, model and DFT energies should be the same). The result is presented in Fig.\ \ref{leastsquares} (b). Similarly, we extract the coupling constants and calculate the energies of the magnetic states using the pure Heisenberg Hamiltonian (in an overdetermined system of equations, the only unknowns are the bilinear coupling constants:  $J_c$, $J_{ab}$, $J_a$, $J_{diag}$, $J_b$ and $J_3$). The model energies plotted versus the energies obtained from first-principles calculations are shown in Fig.\ \ref{leastsquares} (a). Moreover, we extract the deviations of the energies predicted by both Hamiltonians from their values obtained with DFT for each considered magnetic state. These deviations are summarized in the bar charts shown in the insets in Fig.\ \ref{leastsquares}. One can see that the Hamiltonian which includes both bilinear and four-spin terms gives much better agreement with the results of DFT calculations than the pure Heisenberg Hamiltonian. We repeat this analysis also for the Hamiltonian which involves six bilinear exchange couplings and only inter-plane four-spin ring coupling $K$, as $G$ was found to be negligible. The extracted coupling constants as well as the deviations between model and DFT energies remain almost the same as those which were obtained using the full Hamiltonian (which includes also $G$). This means that the addition of just one parameter $K$ into the model Hamiltonian can already significantly improve the description of the magnetism in the considered compound. 

The effect of the higher order exchange interactions in TbMnO$_3$ can also be examined using noncollinear calculations. The simplest approach is to map the angular dependence of the total energy, obtained in Sec.\ \ref{sec3} from the spin rotations from A-AFM to G-AFM orderings, to the Hamiltonian which includes bilinear, biquadratic and four-spin ring interactions by fitting to the function 
\begin{equation}
g(\alpha)=A_2+B_2\cos(\alpha)+C_2\cos^2(\alpha).
\end{equation} 
The result is shown in Fig.\ \ref{fitToHeis} and clearly demonstrates that the introduction of the higher order couplings into the model Hamiltonian greatly improves the fitting. The strengths of these couplings are determined by the fitting parameter $C_2$. Note that this term includes the contributions from four-spin ring exchanges as well as from the in-plane biquadratic couplings and that these terms cannot be separated. Similar behavior of $E(\alpha)$ was found in Ref. \onlinecite{Novak:2008di} for hexagonal YMnO$_3$ using \textit{ab initio} calculations within the LDA+U approximation, where it was discussed only in terms of bilinear and biquadratic exchanges.
\begin{figure}
\centering
{\includegraphics[scale=0.5,trim=1cm 0cm 0cm 0cm]{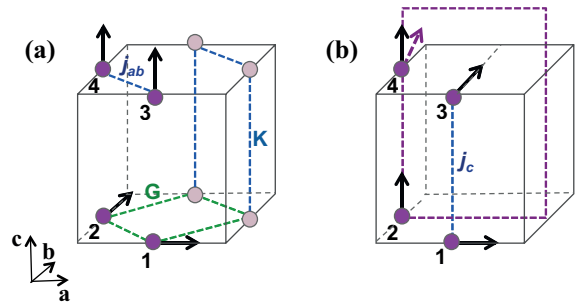}}
\caption
{(Color online) Magnetic orderings which are used to extract the biquadratic exchange interactions: (a) $j_{ab}$ in the $ab$ planes ($G$ and $K$ indicate four-spin ring exchange couplings in plaquettes of Mn spins confined in the $ab$ planes and those containing pairs of spins from neighboring $ab$ planes, respectively); (b) $j_c$ along the $c$ axis (violet dashed rectangle indicates a rotation plane of the spin 4).}  
\label{bqCalc}
\end{figure}
\label{sec5}
\begin{figure*}[!htb]
\begin{minipage}{0.329\linewidth}
\includegraphics[width=1\linewidth,trim=1cm 0cm 0.5cm 0cm]{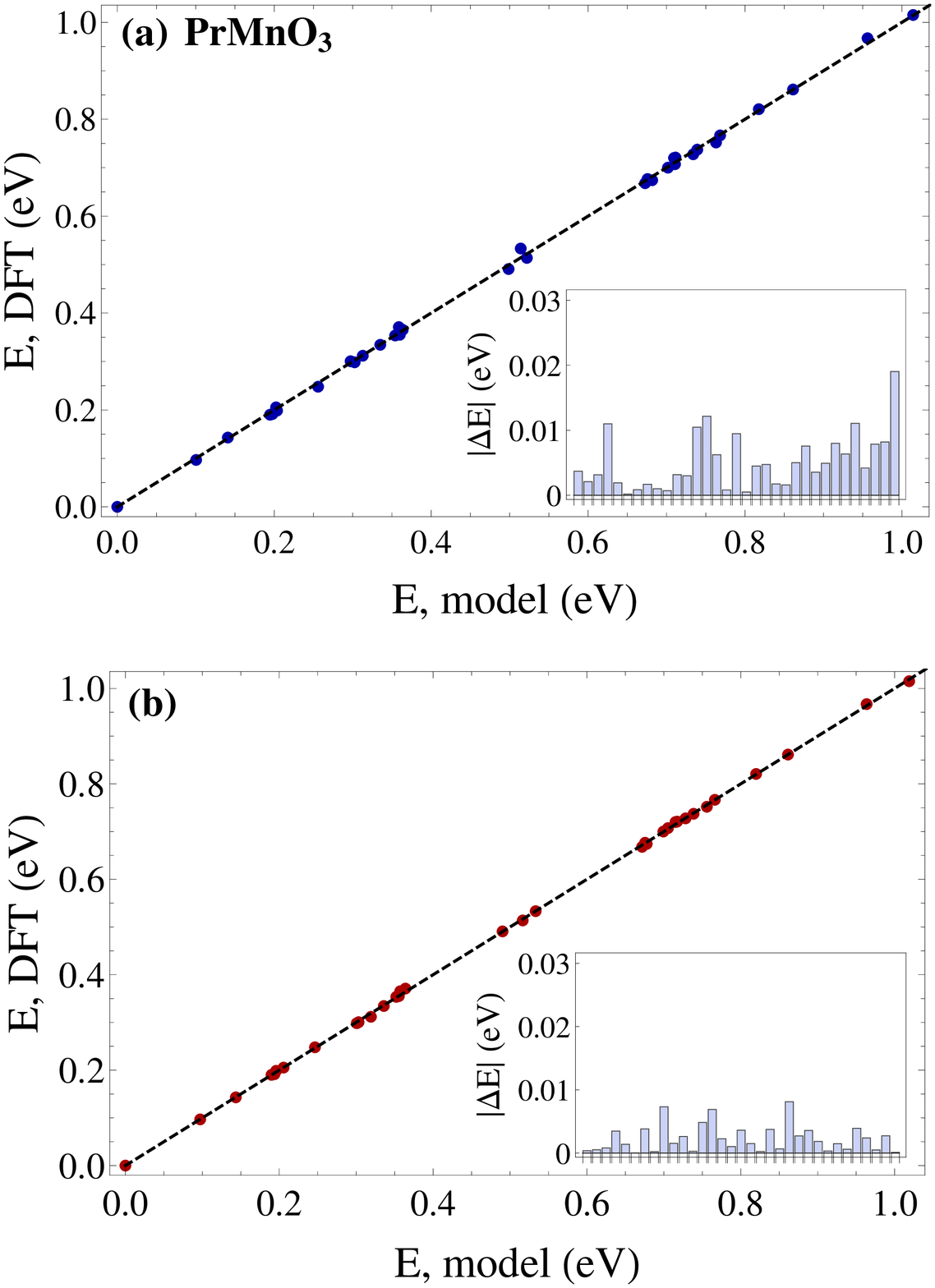} \\
\end{minipage}
\hfill
\begin{minipage}{0.328\linewidth}
\includegraphics[width=1\linewidth,trim=1cm 0cm 0.5cm 0cm]{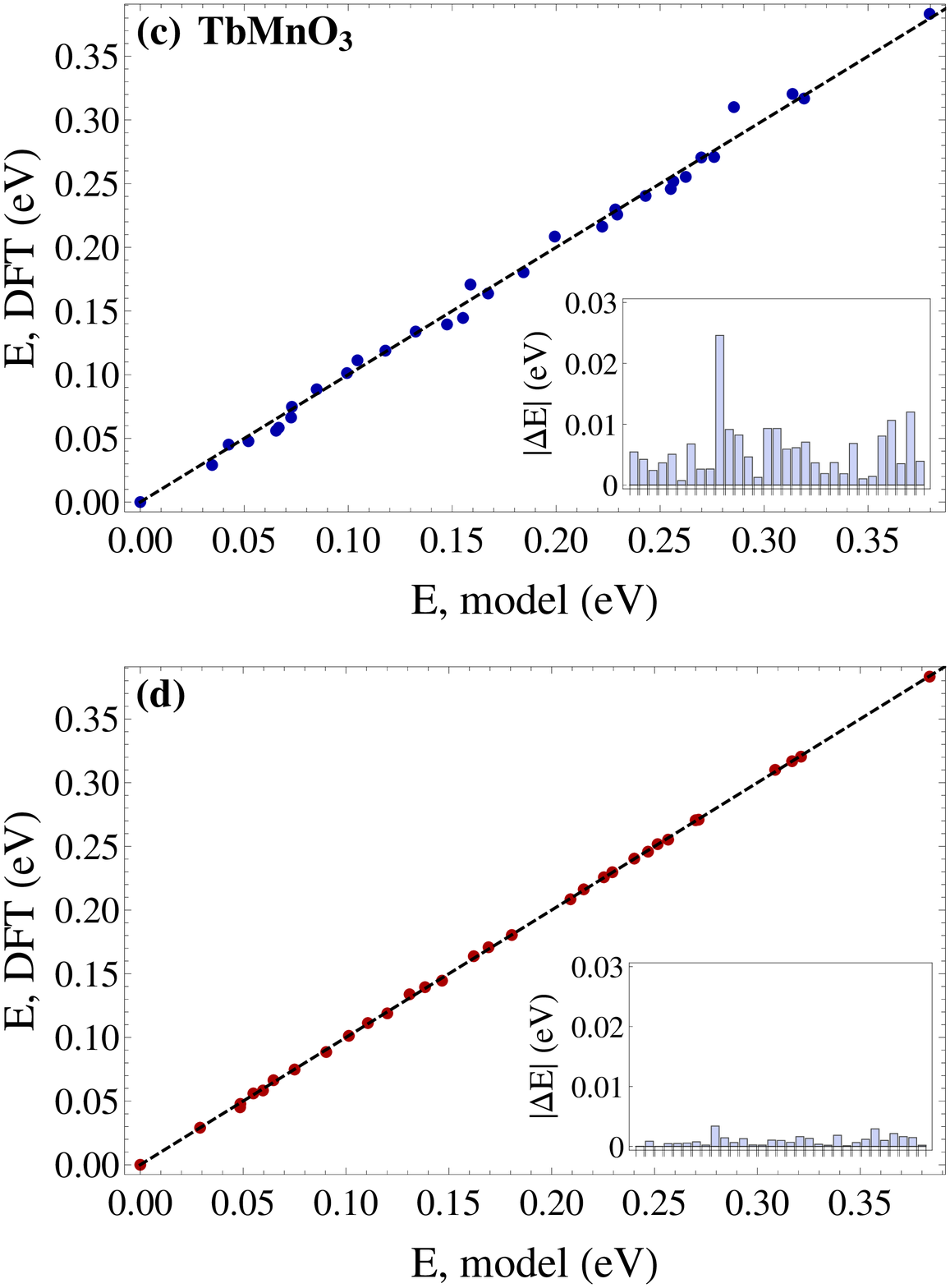} \\ 
\end{minipage}
\hfill
\begin{minipage}{0.329\linewidth}
\includegraphics[width=1\linewidth,trim=1cm 0cm 0.5cm 0cm]{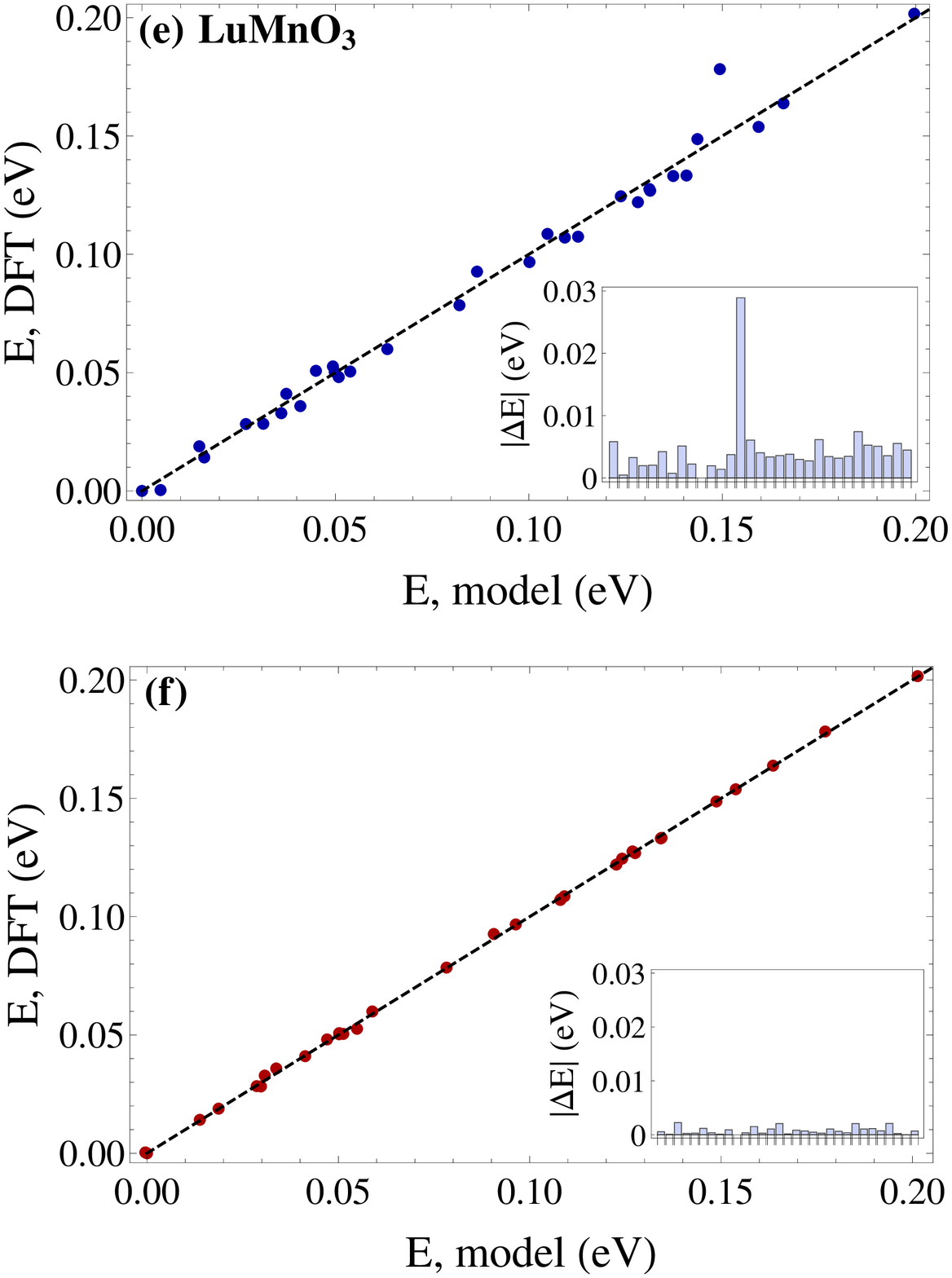} \\ 
\end{minipage}
\caption{(Color online) Energies of the 80 atom supercells of PrMnO$_3$ (a,b), TbMnO$_3$ (c,d) and LuMnO$_3$ (e,f) with more than 30 inequivalent magnetic configurations (referred to that of the lowest-energy state) predicted by the pure Heisenberg Hamiltonian (blue dots) and the Hamiltonian which includes bilinear and four-spin ring exchanges (red dots) and plotted versus the energies of corresponding states calculated using DFT.}
\label{allrmn}
\end{figure*}

To complete the analysis of the full model Hamiltonian we need to estimate the coupling constants which define the biquadratic exchanges in TbMnO$_3$. For this we only take into account the biquadratic interactions between the nearest neighbors in $ab$ planes and along the $c$ axis and denote the corresponding coupling constants as $j_{ab}$ and $j_c$ (see Fig.\ \ref{bqCalc}). The problem can be simplified if we eliminate the contribution from four-spin inter-plane ring exchanges by considering the magnetic states which set to zero at least one scalar product in each of the three terms in Eq.\ \ref{4body}. This can be achieved by setting three Mn spins in the unit cell perpendicular to each other as shown in Fig.\ \ref{bqCalc}. By rotation of the remaining spin the angular dependence of the total energy can be obtained and the coupling constants can be found from the corresponding fittings.
To extract $j_{ab}$ we start from the magnetic state shown in Fig.\ \ref{bqCalc} (a) and rotate spin 4 by an angle $\alpha$ from 0 to $180^\circ$ in the $ac$-plane. The energy of this system can be written as follows:
\begin{eqnarray}
E(\alpha) & = & E+4J_{ab} \cos(\alpha)+8J_{diag} \sin(\alpha)+ \nonumber \\  &+& 4j_{ab}\cos^2(\alpha)+8G\cos^2(\alpha),
\end{eqnarray}
where the third term is given by spins 1 and 4 and all other terms by spins 3 and 4 ($S$=1) and $E$ includes the exchange couplings which are constant at every $\alpha$ for the considered magnetic states and other nonmagnetic interactions. The coupling constants can be extracted by fitting to the function:
\begin{eqnarray}
f(\alpha)&=&A_1+4J_{ab}\cos(\alpha)+8J_{diag}\sin(\alpha)+ \nonumber \\ &+&(D_1+8G)\cos^2(\alpha),
\end{eqnarray}  
where we set the values of $J_{ab}$, $J_{diag}$ and $G$ to those which were extracted in the collinear calculations and presented in the first line of Table \ref{allcouplings}. $D_1/4$ defines $j_{ab}=-2.29$ meV. 
We proceed in a similar way to extract the coupling constant $j_c$. Starting from the magnetic state presented in Fig.\ \ref{bqCalc} (b) and rotating spin 4 by $\alpha$ from 0 to 180$^\circ$ in the $ac$ plane, we obtain $E(\alpha)$. One can see, that in the considered magnetic state neither in-plane nor interplane four-spin ring couplings contribute to $E(\alpha)$ and fitting to
\begin{eqnarray}
g(\alpha)& =& A_2+2J_c \cos(\alpha)+8J_{diag} \sin(\alpha)+\nonumber \\ &+&D_2\cos^2(\alpha),
\end{eqnarray}
gives $j_c=D_2/2=-0.81$ meV. 

Thus we demonstrate that the higher order exchange interactions are significant in TbMnO$_3$ (especially the four-spin ring interplane coupling $K$ and biquadratic in-plane coupling $j_{ab}$) and have to be included in the model Hamiltonian to properly describe the magnetic properties of this material.

\section{Exchange interactions in other o-$R$MnO$_3$}
\label{sec5}
Finally, in this section we investigate the evolution of the exchange couplings in o-$R$MnO$_3$ with increasing GFO distortion due to decrease in the radius of the $R$ cation. For this purpose we consider PrMnO$_3$ and LuMnO$_3$, which have among the largest and the smallest $R$ radii in the series of o-$R$MnO$_3$, respectively. We set $U_\mathrm{Mn}=1$ eV for Mn $d$ states, which gives a correct magnetic ground state for both considered systems. We calculate the total energies of these systems within the 80 atom supercells (experimental unit cells \cite{{Wdowik:2011kz},{Okamoto:2008ey}} are duplicated along $a$ and $b$ directions) for 34 and 32 inequivalent collinear magnetic states, respectively. By writing the expressions for the energies of these magnetic states using the model Hamiltonian that includes bilinear and four-spin ring exchange interactions, we obtain overdetermined systems of equations with respect to six bilinear and two four-spin ring couplings for each compound. In both cases the lowest energy state was taken as the reference (the A-AFM state for PrMnO$_3$ and the E-AFM state for LuMnO$_3$). We solve these systems of equations using the least mean square method and find the values of all coupling constants (Table \ref{allcouplings}). For comparison, we calculate the exchange couplings in a similar way for TbMnO$_3$ for the considered value of $U_\mathrm{Mn}$. 
From Table \ref{allcouplings} one can see that the increasing GFO distortion has the strongest effect on the in-plane NN coupling $J_{ab}$ which changes by more than one order of magnitude from Pr to Lu. It also enhances $J_b$ as we expected (see Sec.\ \ref{sec:2b}). However, the behavior of the other coupling constants (particularly the absence of a trend in the variation of $J_c$ and $J_a$, and the strong change in some coupling constants and weak change in others) with the variation of Mn-O-Mn bond angles still requires a further analysis. 

As the next step, we use the extracted values of the exchange couplings to calculate the energies of all considered magnetic states within this model Hamiltonian and plot them versus the energies of these states which we calculate using DFT (see Fig.\ \ref{allrmn} (b), (d) and (f)). For comparison, we extract similarly the coupling constants using the pure Heisenberg Hamiltonian for each compound, then calculate the energies of all states, predicted by this Hamiltonian, and plot them versus the DFT energies of these states (Fig.\ \ref{allrmn} (a), (c) and (e))). The insets in all resulting graphs show the deviations of the model energy from the DFT energy for each considered state. 

From Fig.\ \ref{allrmn} we find that the model Hamiltonian which includes both the bilinear and four-spin ring exchanges gives much better agreement with the results of the first-principles calculations. One can conclude that the Heisenberg model works relatively well for PrMnO$_3$, since the model and DFT energies almost coincide in Fig.\ \ref{allrmn} (a). However, if one compares the $\Delta E$ values which were obtained for the bilinear-only case (see insets in Fig.\ \ref{allrmn} (a), (c) and (e)), one can see that they are similar and even larger than those of TbMnO$_3$ and LuMnO$_3$, but small relative to the  energy scale of the bilinear couplings, in particular, $J_{ab}$ (see Table \ref{allcouplings}). When GFO distortion increases and $J_{ab}$ drops (as in TbMnO$_3$ and LuMnO$_3$), the $\Delta E$ due to the non-Heisenberg terms become significant. However, when the four-spin ring couplings are added in the model Hamiltonian (Fig.\ \ref{allrmn} (b), (d), (f)), the $\Delta E$ values reduce drastically.

Finally, we extract the biquadratic couplings $j_c$ and $j_{ab}$ for PrMnO$_3$, TbMnO$_3$ and LuMnO$_3$ (see Table \ref{allcouplings}) applying the method which was described in detail at the end of Sec.\ \ref{sec4b}. For all compounds we obtain strong negative in-plane biquadratic couplings $j_{ab}$, which favor collinear alignment of spins within the $ab$ planes and can drive an evolution of a magnetic order from a spiral to an E-AFM state for systems with large GFO distortions. This confirms the finding of Ref.\ \onlinecite{Kaplan:2009fx}, where the biquadratic exchange interaction was claimed to be important in the establishment of the E-AFM order. The in-plane coupling $j_c$ is found to be much more affected by GFO distortion than $j_{ab}$. Again, the origin of this behavior still has to be clarified.

Thus we show that the Heisenberg Hamiltonian cannot accurately predict the magnetic properties of o-$R$MnO$_3$ with large GFO distortions. In these materials the bilinear couplings  become comparable in magnitude with the biquadratic and four-spin ring interactions and it is essential to include the latter two into the model Hamiltonian for proper analysis of the magnetism.

\section{Summary}
\label{sec6}
In summary, we investigated the microscopic exchange couplings in the series of o-$R$MnO$_3$ in order to find an isotropic part of a model Hamiltonian which can properly describe the magnetism in these materials. The work was motivated by the inconsistencies in the results obtained in several theoretical studies when the exchange couplings in o-$R$MnO$_3$ was mapped onto the Heisenberg Hamiltonian as well as by the absence of agreement on the origin of the E-AFM order in o-$R$MnO$_3$ with small $R$ cations ($R$=Ho...Lu). We started our analysis from the most studied multiferroic orthorhombic manganite, TbMnO$_3$, and estimated the exchange couplings with several approaches (collinear and noncollinear) using DFT. We observed a clear deviation from the behavior predicted by the Heisenberg model. Moreover, we confirmed the importance of the AFM third NN in-plane coupling, $J_3$, in the establishment of the spiral state in this compound.  
In the next step we explored whether the observed non-Heisenberg behavior originates from the presence of the orbital ordering in TbMnO$_3$ and its coupling with the Mn spins. To check this, we analyzed the changes in the orbital mixing angle with structural distortions (Jahn-Teller and GFO) and with variation of the magnetic ordering using the Wannier function representation. We found that the orbital mixing angle indeed can be affected by the magnetic order, however, we showed that this change is quite small and is almost unchanged by the structural distortions. In turn, we found that the deviation from Heisenberg behavior does not appear when the amplitude of JT distortion is varied. It appears only with increasing GFO distortion, which decreases the energy of the NN exchange interactions and makes the weak energy contributions more important. We demonstrated, however, that these weak contributions do not originate from the variation of the orbital ordering. Indeed, compounds which do not have an orbital degree of freedom (such as TbCrO$_3$ and TbFeO$_3$) also exhibit a deviation from the energy behavior predicted by the Heisenberg model. Finally, we investigated the effects of exchange couplings of higher order than the ordinary bilinear exchange (biquadratic and four-spin ring interactions), which are usually neglected. We demonstrated that the higher order contributions are significant (especially inter-plane four-spin ring exchange $K$ and biquadratic in-plane coupling $j_{ab}$) and can be comparable with the bilinear exchanges for o-$R$MnO$_3$ with small radii of $R$ cations. We showed that the inconsistent values of the exchange couplings which were obtained from the collinear calculations within the Heisenberg model (Sec.\ \ref{sec3}) can be explained only by addition of the four-spin ring couplings into the model Hamiltonian. Moreover, we proved that such a model Hamiltonian predicts the magnetic properties of o-$R$MnO$_3$ with much higher precision than the pure Heisenberg Hamiltonian. The finding of the strong negative in-plane biquadratic exchange interaction $j_{ab}$, which favors a collinear spin alignment within the $ab$ planes, is in agreement with the suggestion of Ref.\ \onlinecite{Kaplan:2009fx} that $H_{bq}$ is crucial in the establishment of the E-AFM state in o-$R$MnO$_3$ with small radii of the $R$ cations.

\bibliographystyle{apsrev}
\bibliography{paper}

\end{document}